# THz Discrimination of materials: demonstration of a bioinspired apparatus based on metasurfaces selective filters

P. Carelli[1,2], F. Chiarello[2], G.Torrioli[2] and M. G. Castellano[2]

[1]DSFC, Università dell'Aquila, via Vetoio 1, 67100 L'Aquila, Italy
[2]IFN-CNR, via Cineto Romano 42, 00156 Rome, Italy.

fabio.chiarello@ifn.cnr.it
Ph. +39 06 41522 228
Fax. +39 06 41522  220

**Abstract**
We present an apparatus for terahertz fingerprint discrimination of materials designed to be fast, simple, compact and economical in order to be suitable for preliminary on-field analysis. The system working principles, bioinspired by the human vision of colors, are based on the use of microfabricated metamaterials selective filters and of a very compact optics based on metallic ellipsoidal mirrors in air. We experimentally demonstrate the operation of the apparatus in discriminating simple substances such as salt, staple foods and grease in an accurate and reproducible manner. We present the system and the obtained results and discuss issues and possible developments.

## 1. Introduction

The terahertz region of the spectrum (among 0.5 THz and 5 THz [1]), has great importance since it fills up the space of our everyday life in a wide range of disciplines. Despite of its importance it is one of the least explored region of the electromagnetic spectrum mainly because of a series of technological difficulties: low intensity of the known sources, high absorption of the air medium, difficulties in using the techniques developed for radio frequencies and optics at either ends of this range. Despite some developments that occurred in the last century, real improvements have occurred only in the last two decades, in particular thanks to different space born THz observatories (IRAS, ISO, SWAS, Odin, Akari, Herschel) and, on the ground, in the side regions of the THz range (below 1 THz and above 10 THz) [2]. In the last 10 years there have been further developments: the use of QCL (Quantum Cascade Laser) working at THz [3]; the extension of TDS (Time Domain Spectroscopy) and FTIR (Fourier Transform Infrared spectroscopy) to the THz region [2]; the introduction of new superconducting detectors [4]; the utilization of metamaterials, particularly interesting since they allow a strong and controllable coupling of THz radiation with matter [5] and this, for example, allows to implement effective selective filters in the THz region [6].

The use of TDS and FTIR spectroscopy is promising in identification of samples of interest in biology, chemistry and material science, for example: imaging and spectroscopy for security applications (explosives, weapons and drugs) [7], TDS spectroscopy for pharmaceutical development and pharmaceutical process analytical technology [8]; DNA and RNA weakest bond study by FTIR [9]; the dynamical properties of water in the solvation shell of proteins [10]; reflection spectroscopy for medical imaging [11]; study of gases and their mixtures fingerprinted on the basis of the distinct transition frequencies [12]; artists' material analysis (pigment and binders) [13]; detection of antibiotic residues in food [14].

The available instruments are very accurate and can perform precise analysis, but they are not suitable for pre-screening and on-field operations. For example, different intense sources are now available, either monochromatic (QCL) or wideband (femtosecond laser or free electron laser), but at present these sources are unwieldy and not suitable for on-field applications. Moreover, one can easily perform absorption or reflection spectroscopy of almost any material, but it is often necessary a complex preparation of the samples, made in special operative conditions, to avoid artifacts (measurements in vacuum or controlled atmosphere). For these reasons it will be important to develop a simpler, more economical and compact instrumentation (even if less accurate) to be juxtaposed to the more performing one. So we are developing an instrumentation (see ref. [6] for a preliminary approach) that combines existing technologies and new solutions (for sources, filters, detectors etc.) with the final goal to realize a complete system for fast fingerprint identification that must be rapid, compact, economical, simple to use and sturdy enough to make possible on-field analysis. The key point of such instrumentation is the use of metamaterial selective filters that will allow a fingerprint analysis rough but very fast and adequate to the required applications, with a bio-inspired principle based on the human vision of colors. The human eyes are able to distinguish about 10 millions of different colors [15] thanks to just three different kind of cone cells (S, M and L type) which are sensitive to bands, not too narrow, around three separate wavelengths of about 450 nm, 540 nm, 570 nm respectively. Starting from this analogy we propose and demonstrate a fingerprint analysis based on the use of a series of band-pass filters with center frequencies among 1 and 6 THz and with large overlapping bands.

In this work we discuss the latest developments by presenting a complete apparatus and a demonstration of its capability to discriminate different sample substances by a THz transmission analysis.

## 2. Experimental setup

The apparatus is composed of different elements, aligned on the same optical axis (Fig.1): a broadband THz source, a mechanical chopper, a compact focusing optics, a series of interchangeable selective filters based on metasurfaces, and a room temperature detector phase locked to the chopper signal. There are also a system for the mechanical motion, a mechanical support and the electronic interfaces (not shown in figure). In the following paragraphs we will discuss the various parts in detail.

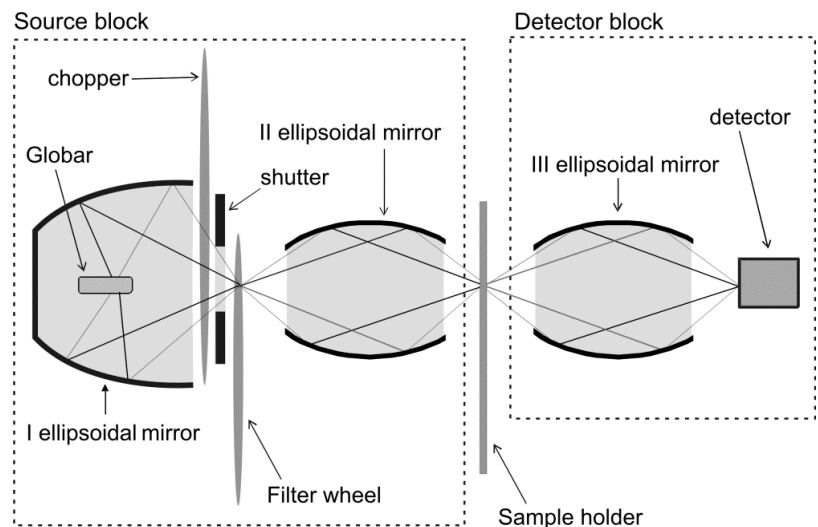

Fig. 1 Scheme of the apparatus.

## 2.1 Source

The chosen source is a broad-band thermal black-body (a globar): a commercial hot surface heater consisting of a silicon carbide rod 5 mm long and 2 mm in diameter [16]. This source is suitable for our purposes and its small dimensions allow a small spot size of the order of millimeters. In the same time it is compact and economical. The power supplied is 20 W and it does not require any cooling. The emission spectrum of this source follows the Planck's law with a temperature of about 1500 K. Note that the THz contribution, even if immersed in a huge infrared background, is not negligible:

$$\frac{\int_{1\,THz}^{5\,THz} B_\nu \, d\nu}{\int_{5\,THz}^{\infty\,THz} B_\nu \, d\nu} \approx 0.011\,\%$$

where $B_\nu$ is the spectral radiance. We have tried also to use a compact plasma source [17] (a metal halide Welch Allyn M21N002 21 W Solarc Lamp), with an higher effective temperature (5000 K) and a larger spectral radiance in the THz region with respect to a hot heater, but unfortunately this kind of source has large power fluctuations on intervals of time of some minutes. Another disadvantage of plasma sources is the large UV and visible background that can cause undesirable effects on the filters, which have a silicon substrate that absorbs the radiation creating holes/electrons and causing a reduction of the transmission.

The beam is modulated by a mechanical chopper placed in front of the source with a typical frequency of 16 Hz. The reference signal from the chopper is used for an accurate lock-in analysis of the detector output.

## 2.2 Focusing optics

We use a focusing optics based on a series of ellipsoidal mirrors with an on-axis configuration (Fig.1) designed in order to reduce the optical path and to collect as much as possible radiation. In particular we use three distinct aluminum mirrors with a overall optical length of 230 mm: the first to collect the radiation from the source and concentrate it on the selective filter; the second to collect radiation from this point and focus it on the sample; the third to focus the radiation on the detector. The physical dimensions of the three mirrors are reported in tab. 1.

|  | Major semiaxis (mm) | Minor semiaxis (mm) | Cut from left focus (mm) | Cut from right focus (mm) |
|---|---|---|---|---|
| Mirror #1 | 35 | 22 | -6.9 | 21.2 |
| Mirror #2 | 50 | 20 | 17 | 8 |
| Mirror #3 | 30 | 10 | 15 | 1 |

Table 1. Dimensions of mirrors. Note that the first mirror is used to concentrate the radiation of the source so that the left focus is inside the mirror. All the other focuses are external to the mirrors' cavities.

Since we are considering radiation with wavelengths between 50 μm and 300 μm we don't need an optical processing of mirrors' surfaces, but a mechanical processing with Computer Numerical Controlled machines, which allows a roughness of about 2 μm on a 1 mm scan, is more than sufficient. For this reason we are able to design and fabricate dedicated optimized mirrors in a short time and at low cost.

An appropriate processing of the mirrors' surfaces can be exploited in order to strongly reduce the undesired background of radiation outside the THz range. We considered and combined together two different approaches: (1) increasing the roughness of the surfaces; (2) covering the surface with an absorbing layer. The roughness of the surface can be modified in a controlled manner in order to diffuse and defocus only the unwanted background radiation and not the THz signals, so that appropriate diaphragms can remove the defocused radiation. Aluminum surfaces can be easily sandblasted, and for this purpose we use an abrasive jet machine with glass beads of around 40 μm diameter producing a roughness of about 8 μm on a 1 mm scan. The rejection of unwanted frequencies can be further enhanced by covering the mirrors' surfaces with a layer of a proper selectively absorbing material. We found that the best choice is a carbon-black layer, simply obtained by burning a paraffin candle on the cold surface of Aluminum. This layer presents an high absorption of infrared and visible radiations and a negligible absorption at THz. In our case the typical thickness of the deposited black carbon layer is around 12 μm with a density lower than 0.1 g/cm$^3$. The thickness measurement, performed with a Scanning Electron Microscope, shows a conformal surface of the layer (a measurement with atomic force microscope or with a surface profilometer was not possible because the black carbon layer is too soft). In Figure 2 the reflectivity of a test flat Aluminum surface with no processing (upper light-gray line) is compared with sandblasting (middle dark-gray line), and with sandblasting and black carbon together (lower black line) in the range 1 THz - 140 THz, measured thanks to a Fourier Transform Infrared Spectrometer (FTIR). The selective effect above 6 THz is noticeable. We tested that the use of all three mirrors with surfaces processed in the same way strongly enhance the out-band rejection (as explained in the following).

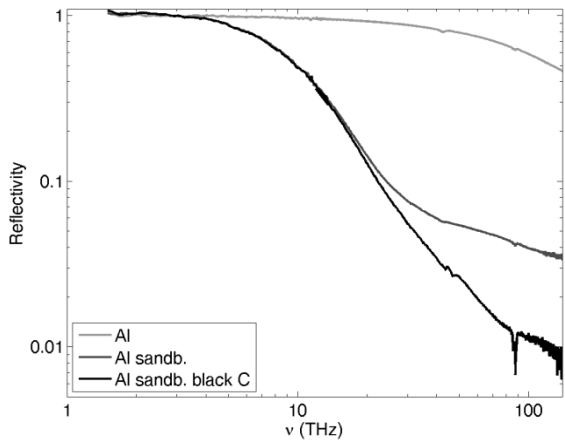

Fig. 2 Reflectivity of Aluminum (upper line), Aluminum sandblasted (middle line), Aluminum sandblasted and covered with black carbon (lower line).

**2.3 Selective filters**

The selective filters allow to identify the signal fingerprints at selected frequency bands. They will be implemented by frequency selective surfaces (FSS), realized by nanofabricated arrays of metallic (aluminum) patterns, with a fabrication process entirely based on Electron beam lithography, similar to that described in ref. [6]. However in this case we improved the technique by placing the same filter on both sides of the substrate, a double polished Si wafer having a diameter of 75 mm, and changing the filters design. A correct electromagnetic design is a key point to have good performances of the band-pass filters. The filters must be adequately selective but not too much, in order to partially overlap and to maximize the collected power. Moreover the effect of sidebands must be strongly reduced. After an optimization based on many electromagnetic simulations we have chosen a periodic complementary annulus shape, which presents the desired features. We have obtained that, for the filter with frequency centered around 3.2 THz, we need a 20 μm side square periodic cell with annulus of 14 μm external diameter and 12 μm internal diameter. Different frequencies are obtained by simply scaling these dimensions. The periodic cell is repeated to cover the filter area (4 mm x 6 mm) on both sides of the wafer substrate. We fabricated a series of 18 rectangular filters with frequencies from 1.12 THz to 5.7 THz, equispaced on the wavelength scale. In addition to the filters there are also two other elements used as reference and for calibration purposes: a not metalized slot (0.8 mm x 6 mm) and a fully metalized region. In this way there are a total of 20 elements placed in equal sectors along the wafer on a circle with mean radius of 25 mm. This configuration allows a fast swapping between filters by simply rotating the wafer by a precise angle, thanks to a step motor.

The filters have been characterized by a FTIR (Fig. 3) obtaining a fairly good agreement between simulated and measured parameters, with merit factors Q approximately constant for all filters and of the order of 3.5. We realized and characterized two identical wafers obtaining a good reproducibility, within 1%.

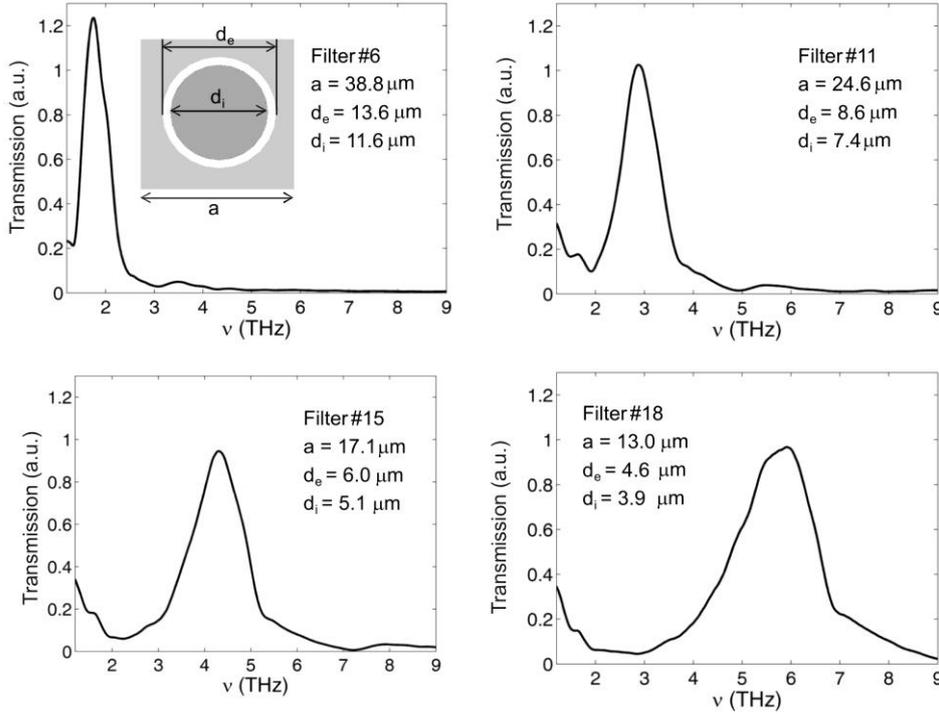

Fig. 3 Transmission of 4 filters measured by means of a FTIR (note that the measurements below 2 THz are affected by a large error due to the FTIR). In the inset of the upper left figure a unitary cell of the filter (Aluminum in gray, not metalized region in white) is represented.

## 2.4 Detector

The used detector is a commercial pyroelectric sensor (QS2-THz-BL Gentec-EO), sensitive to ac signals with a nominal Noise Equivalent Power of 1.5 nW/√Hz, equipped with a Si window (to remove frequencies above the Si gap), mounted inside a box supplied with a battery electronics and preamplified by an external low noise preamplifier. The detector output signal, together with the chopper reference, is digitized and acquired by a computer system that performs a software real time lock-in analysis. The software normalizes the trigger reference to -1/+1 and generates a second reference shifted by a phase of 90° respect to the first one. Then it multiplies these two references by the preamplified pyroelectric output and integrates the results on a fixed number of periods to obtain the components X and Y, the amplitude $A = \sqrt{X^2 + Y^2}$ and the phase $\varphi = \arctan\left(\frac{Y}{X}\right)$.

## 2.5 Apparatus
Fig.4 presents a photograph of the final apparatus. From the left, inside the metallic box, there are the source, the first collecting mirror and the chopper.

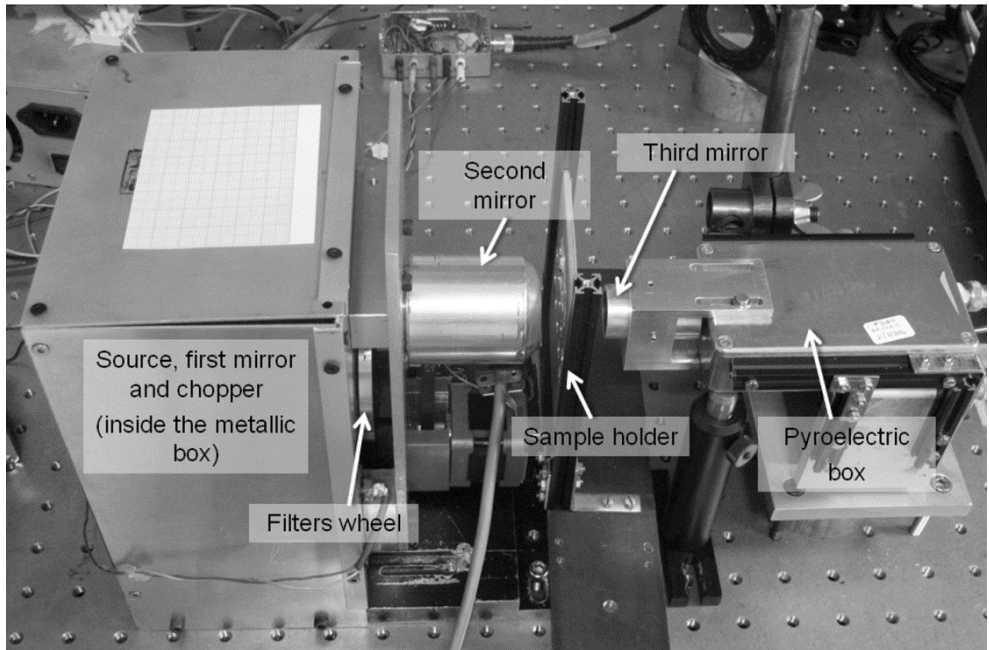

Fig. 4 Photograph of the apparatus.

The source is in correspondence of the left focus of the first mirror while the right focus is outside the box, beyond a hole in the box wall acting as first diaphragm. Inside the box, at a few millimeters between the end of the mirror and the hole, there is also a flat metallic chopper with a sequence of 4 alternated empty/full sectors (8 sectors overall), rotating at a typical frequency of 4 Hz (corresponding to a reference signal of about 16 Hz). Just outside the box there is an aluminum holder (wheel) containing the wafer with the filters. The wheel presents 20 holes in correspondence of the 20 elements on the wafer. It can be rotated with precision (0.9° per step) to select the desired filter, thanks to a computer controlled step motor. The wheel is positioned in order to have the chosen filter exactly in correspondence of the right focus of the first mirror and of the left focus of the second mirror, which is fixed to the external part of the box. On the right focus of the second mirror it is placed the sample holder that can be moved with precision, thanks to a computer controlled xyz translation stage (not visible in Fig.4). A third mirror is fixed on the pyroelectric box, with the left focus on the sample holder and the right focus on the pyroelectric sensor. All the mirrors are sandblasted and covered with black-carbon. A computer system, based on LabVIEW, controls the 4 step motors (x, y, z and filters rotation) through an Arduino board [18] specially programmed, and at the same time acquires the pyroelectric signal and the chopper reference and performs the software lock-in analysis previously described.

In an initial phase the different elements of the system were mounted and aligned separately on an optical bench. In order to arrive to our final configuration, with all the elements assembled in a single block, we used a stepwise procedure. We started from a few elements (source, first mirror and chopper) and we found the focus by moving the detector thanks to the xyz stage. Then we added the metallic box, the wheel and the second mirror and adjusted their positions manually by maximizing the signal detected by the sensor, moved again by the xyz stage. Once fixed the position of the previous elements (second mirror and wheel) we added the third mirror and optimized its position with a similar procedure. Afterwards the third mirror was fixed to the pyroelectric box. At this point we have two independent blocks: the source block with the first two mirrors, the chopper and the wheel, and the detector block with its third mirror. The source block can be used as a stand-alone quite-monochromatic THz source with changeable frequency. The detector block can be used as a stand-alone THz imaging sensor, thanks to its small spatial resolution and to the xyz movement stage. We measured the spatial resolution of the complete apparatus (given by the dimensions of the focus spot on the sample position) by moving the detector block respect to the source block. Eventually the two blocks were fixed together in the optimal relative position and the xyz stage is now used to move the sample holder inside the focus spot.

Thanks to the short optical path we are able to work in air, but it is possible to improve the system by enclosing it in a controlled atmosphere.

### 3. Measurements

As previously described it is possible to evaluate the spatial resolution of the apparatus by moving the detecting block (consisting of the detector box plus the third mirror) respect to the source block thanks to the xyz stage. This allows to measure the dimension of the focus spot where the sample will be placed. In fig.5 the map obtained with no sample and filters on the optical path (open slot in the filters wheel) is plotted. It is obtained by moving the detector block in the directions transverse to the beam (yz) for fixed (and optimized) x.

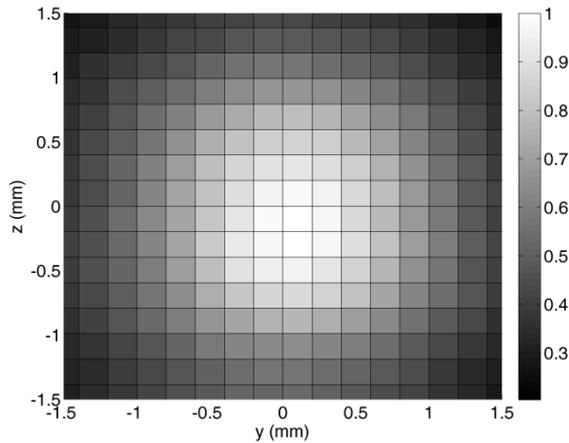

Fig. 5 Map of the focus spot (directions yz transverse to the beam).

The measurement gives a typical spot size in the yz (transverse) directions of about 2.1 mm, obtained as the FWHM of the Gaussian fitting data. The spot size in the x (longitudinal) direction is obtained by maintaining the optimal yz position and moving the block in the x direction. We measured about 9 mm.

Afterwards, in order to measure the transmission characteristic of the different filters, we fixed together the two blocks in the optimal position and rotated the filter wheel in the absence of samples. In Fig.6 the measured signal for different angles of rotation of the filters wheel is plotted. The first observed peak (on the left) corresponds to the open slot, followed by the fully metalized region (plateau around 30°) and by the 18 filters in decreasing order of frequency. The decreasing of peaks height is due to the Rayleigh-Jeans law, as expected. The minima correspond to the positions between adjacent filters, where metal partially covers the beam. The background noise has been measured by covering the source completely. The result is about 1.6 nW/√Hz, close to the nominal NEP of the detector.

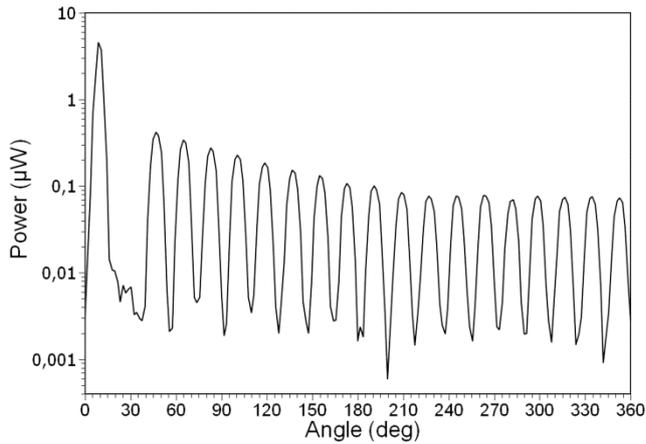
Fig. 6 Transmitted power for different angles of rotation of the filters wheel.

The measurement of the peaks height is an indication of the correct operation of the apparatus. This measurement has been previously done with mirrors not processed by sandblasting and black-carbon covering. In that case the peaks presented almost identical heights, an indication that the signal was dominated by a large, not removed infrared background. In that condition the substance analysis was not possible. By using treated mirrors surfaces the problem has been removed, and this gives an indirect indication concerning the effectiveness of the treatment.

Our final goal is to demonstrate the capability of using this system to distinguish different sample materials. For this purpose we acquired the transmitted power of all the 18 filters for different samples, swapped on the focus spot thanks to the xyz stage moving the sample holder. We have chosen three distinct categories of materials: grease, in particular Apiezon (A) (used also as a support for other substances); alkali halide salt, in particular Potassium Bromide KBr (E); staple foods, in particular pastry flour (B), durum wheat (C), yeast (D) and brewer yeast (F). The substances have been prepared as fine powders (except the grease) and fixed on a 10 μm High Density Polyethylene (HDPE) film thanks to a thin layer of Apiezon grease. The chosen measurement time for each single filter was 10 s, so that the complete characterization for a single substance with all the 18 filters required 3 minutes. By considering the NEP of 1.6 nW/√Hz, this time corresponds to an error of the order of a few nW. Each characterization was repeated more times in order to test the repeatability and the obtained result is within the evaluated error. We analyzed the results by measuring the power transmitted through a substance for each filter and dividing this for the reference power, that is the corresponding signal in the absence of sample. The obtained relative transmittances for each substance is plotted in Fig.7a. Note that these are only "pseudo-spectra", that is the convolution of the transmittance spectra for the different substances with the transmittance of filters. The horizontal positions of symbols in Fig. 7a correspond to the central frequency of each filter, the dimension of these symbols give the typical magnitude of the measurement error (the connecting lines are only for visual aid).

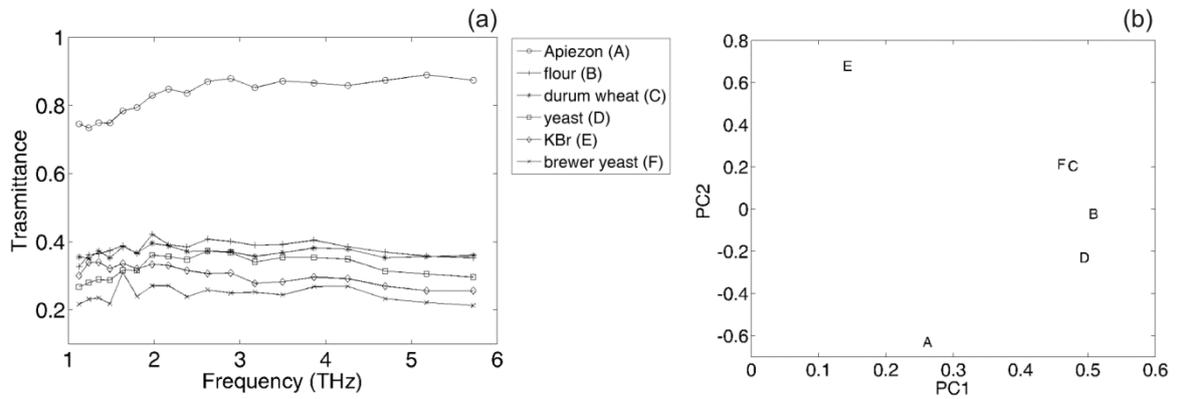

Fig. 7 (a) Measured relative transmittance of the six sample substances (different curves) for all the 18 filters (simbols). (b) Principal Component Analysis scatterplot obtained for the same data reported on the left.

In the pseudo-spectra of Fig. 7a there are (hardly) visible some specific fingerprints of the different groups of substances. In order to distinguish these substances more clearly we use the Principal Component Analysis [19,20] (PCA), a multivariate statistics technique widely used, in particular for classification problems of this kind. It is essentially based on the identification of a linear transformation of the problem from the set of original parameters (the relative transmittances corresponding to the 18 filters) to a new reduced set of the most significant parameters, called Principal Components (PC). In Fig.7b the data reported in Fig.7a are plotted as a scatterplot in functions of just two Principal Components (PC1 and PC2) obtained with the PCA technique. The different typologies (grease, staple foods and salt) are clearly distinguishable in well separated groups and, inside the staple foods category, a more detailed distinction is also visible.

The PCA allows also to reduce the measurement time by carefully choosing only a reduced set of filters, still allowing an adequate material discrimination, even if with less accuracy. Just to give an example, a set of only 6 filters instead of 18 (Fig.8a) reduces the time for a single pseudo-spectrum acquisition from three minutes to a single minute, while the obtained PCA scatterplot (Fig. 8b) still allows a clear distinction between different categories, even if the distinction between flour (B) and durum wheat (C) becomes blurred.

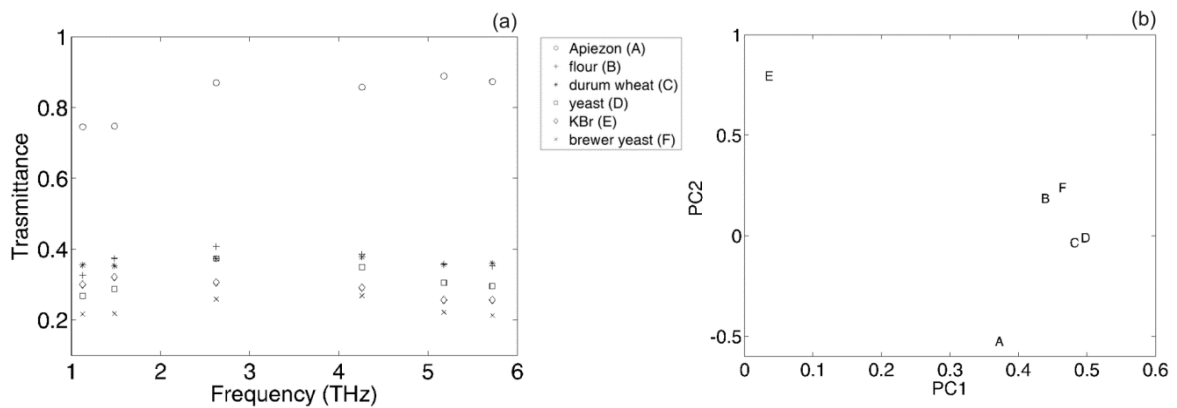

Fig. 8 (a) Measured relative transmittance of the six sample substances (different curves) for a reduced set of 6 filters (carefully chosen thanks to the PCA). (b) Principal Component Analysis scatterplot for the same data.

## 4 Conclusions

We presented an apparatus for the THz discrimination of substances and successfully demonstrated its operation for a few solid samples. The apparatus is based on the bio-inspired principle of the human vision of colors, transposed to the THz region with the use of selective filters based on metamaterials. The apparatus results particularly compact, robust, economical, simple and suitable for on-field applications.

For the next future we want to test the apparatus with a larger set of samples and to optimize the entire system. For example, by rotating the entire apparatus in a vertical configuration instead of the present horizontal one, it will be possible to place the samples substance on an horizontal plane instead of on a vertical one. This will probably simplify and extend the possibility to characterize different sample substances, for example low viscosity liquids or powder without the use of a fixative. Moreover we want to modify the apparatus in order to realize reflection measurements. This will be possible by introducing off-axis rotation of the source block respect to the detector block. A further possibility is the use of the system for "colored" imaging of samples.

**Acknowledgments**
The presented work was partially supported by the Italian Defense Ministry under the project P.N.R.M. n. a2012.169, and by Crisel Instruments. We thank Eugenia Finocchiaro and Carlo Florio for useful discussions, Alessandro Nucara for practical support, Francesca Bertani for useful suggestions and Anna Marletta.